\documentclass[11pt]{article}
\def\directunion{\hbox{$\bigcirc$ \hskip - 11.3 pt \raise 0.1pt
\hbox{$\scriptstyle \vee$}}\ }

\newtheorem{theorem}{Theorem}
\newtheorem{definition}{Definition}

\newtheorem{proposition}{Proposition}
\newtheorem{axiom}{Axiom}

\usepackage[psamsfonts]{amssymb}
\newcommand{\compl}{{\mathbb C}}

\newcommand{\bd}{\begin{definition}}
\newcommand{\ed}{\end{definition}}
\newcommand{\bp}{\begin{proposition}}
\newcommand{\be}{\begin{equation}}
\newcommand{\ee}{\end{equation}}
\newcommand{\bea}{\begin{eqnarray}}
\newcommand{\eea}{\end{eqnarray}}
\newcommand{\ba}{\begin{array}}
\newcommand{\ea}{\end{array}}
\begin{document}

\title{The Linearity of Quantum Mechanics at Stake: The Description of
Separated Quantum Entities\footnote{Published as: D. Aerts and F. Valckenborgh, ``The linearity of quantum
mechanics at stake: the description of separated quantum entities", in {\it Probing the Structure of Quantum
Mechanics: Nonlinearity, Nonlocality, Computation and Axiomatics}, eds. D. Aerts, M. Czachor and T. Durt,
World Scientific, Singapore (2002).}}

\author{Diederik Aerts and Frank Valckenborgh}
\date{}
\maketitle

\centerline{Center Leo Apostel (CLEA) and}
\centerline{Foundations of the Exact Sciences
(FUND),}
\centerline{Brussels Free
University, Krijgskundestraat 33,}
\centerline{1160 Brussels,
Belgium.}
\centerline{diraerts@vub.ac.be, fvalcken@vub.ac.be}

\begin{abstract}
\noindent
We consider the situation of a physical entity that is the
compound entity consisting of two
`separated' quantum entities. In earlier work it has been proved by one of
the authors that such a
physical entity cannot be described by standard quantum mechanics. More
precisely, it was shown that two of the axioms of traditional
quantum axiomatics are at the
origin of the impossibility for
standard quantum mechanics to describe this type of compound entity. 
One of these
axioms is equivalent with the superposition principle, which means
that separated quantum entities
put the linearity of quantum mechanics at stake. We analyze the conceptual
steps that are involved in this
proof, and expose the necessary material of quantum axiomatics to be able
to understand the argument.
\end{abstract}

\section{Introduction}
It is often stated that quantum mechanics is basically a linear theory. Let us
reflect somewhat about what one usually means when expressing this
statement.

The Schr\"odinger equation that describes the change of the state of a
quantum entity under the influence of the
external world is a linear equation. This means that if the wave function
$\psi_1(x)$ and the wave function
$\psi_2(x)$ are both solutions of the Schr\"odinger equation, then, for
$\lambda_1, \lambda_2 \in \compl$, the wave function
$\lambda_1 \psi_1(x) +
\lambda_2\psi_2(x)$ is also a solution. Hence the set of solutions of the
Schr\"odinger equation forms
a vector space over the field of complex numbers: solutions can be added and
multiplied by a complex number and the results remain
solutions. This is the way how the linearity of the evolution equation is
linked to the linearity, or vector space structure, of
the set of states.

There is another type of
`change of state' in quantum mechanics, namely the one provoked by a
measurement or an experiment. This type of
change, often called the `collapse of the wave function', is nonlinear. It
is described by the action of a
projection operator associated with the self-adjoint operator that
represents the
considered measurement -- and hence for this part it is
linear, because a projection operator is a linear function -- followed by a
renormalization of the state. The two effects,
projection and renormalization, one after the other, give rise to a
nonlinear transformation.

The fundamental nature of the linearity of the vector space used to
represent the
states of a quantum mechanical entity is expressed by adopting the
`superposition principle' as one of the basic principles of
quantum mechanics. Because linearity in general appears very often as an
idealized case of
the real situation, some suspicion towards the fundamental linear nature of
quantum mechanics is at its place. Would there not be
a more general theory that as a first order linear approximation gives rise
to quantum mechanics?

In relation with this question it is good to know that the situation in
quantum mechanics is very different from the situation in
classical physics. Nonlinear situations in classical mechanics exist at all
places, and it can easily be understood how
the linearized version of the theory is an idealization of the general
situations (e.g. the linearization used to study small
movements around an equilibrium position). Quantum mechanics on the
contrary was immediately formulated as a linear theory, and
no nonlinear version of quantum mechanics has ever been proposed in a
general way.

The fundamentally different way in which linearity presents itself in quantum
mechanics as compared to
classical mechanics makes that quite a few physicists believe that quantum
linearity is a profound property of the world. The way
in which classical mechanics works as a theory for the macroworld with
at a `more basic level' quantum mechanics as a description for the
microworld, and additionally the hypothesis that this macroworld is built
from building blocks that are quantum, makes
that some physicists propose that the nonlinearity of macrophenomena should
emerge from an underlying
linearity of the microworld. This type of reflections is however very
speculative. Mostly because nobody has been able to
solve in a satisfactory way the problem of the
classical limit,
and explain how the microworld described by quantum mechanics gives rise to
a macroworld described by classical mechanics.

Because of the profound and unsolved nature of this problem it is worth to
analyze a result that has been obtained by
one of the authors in the eighties. The result is the
following:

\begin{quotation}
\noindent {\it If we consider the physical entity that consists of two
`separated' quantum entities, then this physical entity
cannot be described by standard quantum mechanics \cite{Aerts1981,Aerts1982}.}
\end{quotation}
The aspect of this result that we want to focus on in this article, is that
the origin of the impossibility for standard quantum
mechanics to describe the entity consisting of two separated quantum
entities is the linearity of the vector space
representing the states of a quantum entity.

We analyze the conceptual steps to arrive at this result in the present
paper without giving proofs. For
the proofs we refer to \cite{Aerts1981,Aerts1982}.

\section{Quantum Axiomatics} \label{sec:quantumaxiomatics}
As we mentioned in the introduction, there is no straightforward way to
conceive of a more general, possibly nonlinear,
quantum mechanics if one starts conceptually from the standard quantum
mechanical formalism. The reason is that standard
quantum mechanics is elaborated completely around the vector space
structure of the set of states of a quantum entity and the
linear operator algebra on this vector space. If one tries to drop
linearity starting from this structure
one is left with nothing that remains mathematically relevant to work with.

We also mentioned that there is one transformation in standard quantum
mechanics that is nonlinear, namely the transformation of
a state under influence of a measurement. The nonlinearity here comes from
the fact that also a renormalization procedure is
involved, because states of a quantum entity are not represented by
vectors, but by normalized
vectors of the vector space. This fact gives us a first hint of where to
look for possible ways to generalize quantum mechanics
and free it from its very strict vector space strait jacket. This is also
the way things have happened historically.
Physicists and mathematicians noticed that the requirement of normalization
and renormalization after projection means
that quantum states `live' in the projective geometry corresponding with the
vector space. The standard quantum
mechanical representation theory of groups makes full use of this insight:
group representations are projective
representations and not vector space representations, and experimental
results confirm completely that it is the projective
representations that are at work in the reality of the microworld and not
the vector space representations.

Of course, there is a deep mathematical connection
between a projective geometry and a vector space, through what is called
the `fundamental representation theorem of projective
geometry' \cite{Artin1957}. This theorem states that every projective
geometry of dimension greater than two can be represented
in a vector space over a division ring, where a ray of the vector space
corresponds to a point of the projective geometry, and a
plane through two different rays corresponds to a projective line.
This means that a projective geometry entails the type of linearity that is
encountered in quantum mechanics. Conceptually however a projective
geometric structure is quite different from a vector space
structure. The aspects of a projective geometry that give rise to linearity
can perhaps more easily be generalized than this
is the case for the aspects of a vector space related to linearity.

John von Neumann gave the first abstract mathematical formulation of
quantum mechanics \cite{vonNeumann1932}, and proposed an
abstract complex Hilbert space as the basic mathematical structure to
work with for
quantum mechanics. If we refer to standard
quantum mechanics we mean quantum mechanics as formulated in the seminal
book of von Neumann. Some years later he wrote an important
paper, together with Garrett Birkhoff, that initiated the research on quantum
axiomatics \cite{BvN1936}. In this paper Birkhoff and
von Neumann propose to concentrate on the set of closed subspaces of the
complex Hilbert space as the basic mathematical
structure for the development of a quantum axiomatics. In later years
George Mackey wrote an influential book on the mathematical
foundations of quantum mechanics where he states explicitly that a physical
foundation
for the complex Hilbert structure should be looked for \cite{Mackey1963}. A
breakthrough came with the work of Constantin Piron
when he proved a fundamental representation theorem \cite{Piron1964}. It
had been noticed meanwhile that the set of closed
subspaces of a complex Hilbert space forms a complete, atomistic,
orthocomplemented lattice
and Piron proved the converse, namely that a complete, atomistic
orthocomplemented lattice, satisfying some extra conditions,
could always be represented as the set of closed subspaces of a generalized
Hilbert space \cite{Piron1964,Piron1976}. In his
proof Piron first derives a projective geometry and then makes the step to
the vector space. Piron's representation theorem is
exposed in detail in theorem \ref{theor:representation02} of the present
article.

As we will see, it is exactly the extra conditions, needed to represent
the lattice as the lattice of closed subspaces of a
generalized Hilbert space, that are not satisfied for the description of
the compound entity that consists
of two separated quantum entities. Since the aim of this article is to put
forward the conceptual
steps that are involved in the failure of standard
quantum mechanics to describe such an entity, we will start by
explaining the
general aspects of quantum axiomatics in some detail, omitting all proofs,
for
the sake of readability. For the reader who is interested in a
more detailed
exposition, references to the literature are given.

\subsection{What Is a Complete Lattice?}

A lattice ${\cal L}$ is a set that is equipped with a partial order
relation $<$. This means a relation such that for $a, b, c
\in {\cal L}$ we have:
\bea
a &<& a \label{eq:reflexive} \\
a < b\ &{\rm and}&\ b < a \Rightarrow a = b \label{eq:antisymmetric}\\
a < b\ &{\rm and}&\ b < c \Rightarrow a < c \label{eq:transitive}
\eea
(\ref{eq:reflexive}) is called reflexivity, (\ref{eq:antisymmetric}) is
called antisymmetry and
(\ref{eq:transitive}) is called transitivity of the relation $<$. Hence a
partial order relation is a reflexive, antisymmetric and
transitive relation.

Let us give two examples of partial order relations.
First, consider a set
$\Omega$ and let
${\cal P}(\Omega)$ be the set of all subsets of $\Omega$. The
inclusion operation
$\subset$ on ${\cal P}(\Omega)$ is a partial order on ${\cal P}(\Omega)$.
Second, consider a complex Hilbert space ${\cal H}$, and let ${\cal
P}({\cal H})$ be the set of all closed subspaces of ${\cal H}$. The inclusion
operation $\subset$ on ${\cal P}({\cal H})$ is a
partial order on ${\cal P}({\cal H})$. The two examples that we propose
here are the archetypical examples for quantum
axiomatics. The first example represents (part of) the
mathematical structure related to a classical physical
entity, where $\Omega$ corresponds with its state space, and the second example
the one of a quantum physical entity, where ${\cal H}$ is the complex Hilbert
space representing the states of the quantum entity.

A lattice is a mathematical object that has some more structure than just this
partial order relation. Suppose we consider two elements
$a, b \in {\cal L}$ of the lattice, then we demand that there exists an
infimum and a supremum for $a$ and $b$ for the partial
order relation
$<$  in ${\cal L}$. The infimum and supremum are denoted respectively
$a \wedge b$ and $a \vee b$. An infimum of $a$ and $b$ is a greatest lower
bound, this means a
maximum of all the elements of ${\cal L}$ that are smaller than $a$ and
smaller than $b$. A supremum is a least upper bound,
this means a minimum of all the elements of ${\cal L}$ that are greater
than $a$ and greater than $b$. Let us repeat this in a
formal way. For $a, b \in {\cal L}$ there exist $a \wedge b, a \vee b \in
{\cal L}$ such that for $x, y \in {\cal L}$
we have:
\bea
x < a\ {\rm and}\ x < b &\Leftrightarrow& x < a \wedge b \label{eq:infimum} \\
a < y\ {\rm and}\ b < y &\Leftrightarrow& a \vee b < y \label{eq:supremum}
\eea
A structure $({\cal L}, <, \wedge, \vee)$, where ${\cal L}$ is a set, $<$
is a partial order relation satisfying
(\ref{eq:reflexive}), (\ref{eq:antisymmetric}) and (\ref{eq:transitive}), and
$\wedge$ and
$\vee$ are an infimum and a supremum, satisfying (\ref{eq:infimum}) and
(\ref{eq:supremum}) is a {\bf lattice}.

Our two previous examples are both examples of lattices. For the
case of ${\cal P}(\Omega)$ the infimum and supremum of two
subsets $A, B \in {\cal P}(\Omega)$ are given respectively by the
intersection $A \cap B \in {\cal P}(\Omega)$ and the union $A
\cup B \in {\cal P}(\Omega)$ of subsets. For the case of ${\cal P}({\cal
H})$ the infimum of two closed subspaces $A, B \in
{\cal P}({\cal H})$ is given by the intersection $A \cap B \in {\cal
P}({\cal H})$. On the other hand, the union of two closed subspaces is in
general not a closed subspace. This means that the union does not give us
the supremum in this case. For two closed subspaces $A, B \in {\cal
P}({\cal H})$, the smallest closed subspace that contains both, is
$\overline{A+B}$, the topological closure of the sum of the
two subspaces. Hence this is the supremum of $A$ and $B$ in ${\cal P}({\cal
H})$.

It is worth remarking that we can see already at this level of the 
discussion one
of the fundamental differences between a classical entity and a
quantum entity. The vectors that are contained in the topological closure
$\overline{A+B}$ of the sum of $A$ and
$B$ are exactly the vectors that are superpositions of vectors in $A$ and
vectors in $B$. Hence for a quantum entity, described
by
${\cal P}({\cal H})$, there are additional vectors in the supremum of $A$
and $B$, contained neither in $A$ nor
in $B$, while for the classical entity, described by ${\cal
P}(\Omega)$, there are no such additional elements, because
the supremum of $A$ and $B$ is the union $A \cup B$. This means that by
changing our description to the level of the
lattice, we can express both cases, the classical case and the quantum
case, in the same mathematical language, which is not true
if we describe the quantum entity in a Hilbert space and the classical
entity in a state (or phase) space.

A {\bf complete lattice} is a lattice that contains an infimum and a
supremum for every subset of its elements, hence not only for
each pair of elements as it is the case for a lattice. Let us express
this requirement in a formal way. For $A \subset
{\cal L}$ there exists $\wedge A$ and $\vee A$ such that:
\bea
x < a\ \forall a \in A &\Leftrightarrow& x < \wedge A
\label{eq:completeinfimum} \\
a < y\ \forall a \in A &\Leftrightarrow& \vee A < y \label{eq:completesupremum}
\eea
Note that a complete lattice ${\cal L}$ contains always a minimal
element, namely the element $\wedge{\cal L}$, that we denote
by $0$, and a maximal element, namely $\vee{\cal L}$, that we denote by $I$.

Our two examples, $({\cal P}(\Omega), \subset,
\cap, \cup)$ and $({\cal P}({\cal H}), \subset, \cap,
\overline{\ +\ })$ are both complete lattices. The minimal elements of
${\cal P}(\Omega)$ and ${\cal P}({\cal H})$ are respectively $\emptyset$
and $\{0\}$, and the maximal elements are $\Omega$ and ${\cal
H}$.

In the axiomatic approach the elements of the lattice represent the {\bf
properties} of the physical entity under study. Let us explain how the
states of a physical entity are
represented in the axiomatic approach.

\subsection{Atoms of a Lattice}
An element of a (complete) lattice is called an atom, if it is a smallest
element different from $0$. Let us define
precisely what we mean. We say that $p \in {\cal L}$ is an atom of ${\cal
L}$ if for $x \in {\cal L}$ we have:
\be
0 < x < p \Rightarrow x = 0\ {\rm or}\ x = p
\ee
The atoms of  $({\cal P}(\Omega), \subset, \cap, \cup)$ are the singletons
of the phase space $\Omega$, and the atoms of $({\cal
P}({\cal H}),
\subset,
\cap,
\overline{\ +\ })$ are the one
dimensional subspaces (rays) of the Hilbert space ${\cal H}$.

In the traditional versions of the axiomatic approach the states of the
physical entity under study were
represented by the atoms of the lattice ${\cal L}$. Of course, this is
largely due to the fact that for the
case of standard quantum mechanics, the rays can indeed be identified as
atoms of the lattice of closed
subspaces of the Hilbert space. On the other hand, from a physical point of
view, it is obvious that a state
is a completely different concept than a property, and hence for an
operational approach states should not be
properties. Already in \cite{Aerts1981,Aerts1982} it can be seen that when
it comes to calculating and
proving theorems, the states are treated differently from the properties,
although the old tradition of
representing both within the same mathematical structure, reminiscent of the
`states are represented by atoms
of the lattice' idea, is maintained in \cite{Aerts1981,Aerts1982}. Only
slowly the insight grew about how to
handle this problem in a more profound way. The new way, that is fully
operationally founded, is to introduce
two sets from the start, the set of {\bf states} of the physical entity,
denoted by $\Sigma$, and the set of
properties, denoted by ${\cal L}$. It follows from the operational part of
the construction that additionally
to these two sets one needs to consider a one-to-one function $\kappa: {\cal L}
\rightarrow {\cal P}(\Sigma)$, called
the Cartan map, such that $p \in \kappa(a)$ expresses the following
fundamental physical situation: ``The
property $a \in {\cal L}$ is `actual' if the physical entity is in state $p
\in \Sigma$". Moreover it follows
from the operational aspects of the axiomatic approach that $\kappa$
satisfies the three following additional
requirements:
\bea
\kappa(\wedge_ia_i) &=& \cap_i \kappa(a_i) \label{eq:cartan01}\\
\kappa(0) &=& \emptyset \label{eq:cartan02} \\
\kappa(I) &=& \Sigma \label{eq:cartan03}
\eea
for $(a_i)_i \in {\cal L}$ and $0$ and $I$ being respectively the minimal
and maximal element of ${\cal
L}$. The triple $(\Sigma, {\cal L}, \kappa)$, where $\Sigma$ is a set,
${\cal L}$ a complete lattice, and
$\kappa$ a map satisfying (\ref{eq:cartan01}), (\ref{eq:cartan02}), and
(\ref{eq:cartan03}) has been called a
state-property system \cite{Aerts1999,ACVV1999}, and it is this
mathematical structure that can be
derived from the operational aspects of the axiomatic approach.

Let us consider our two examples and see how this structure appears there.
For the quantum case, $\Sigma$ is
the set of rays of the Hilbert space and ${\cal L}$ the set of closed
subspaces. The Cartan map maps each
closed subspace on the set of rays that are contained in this closed
subspace, and indeed
(\ref{eq:cartan01}), (\ref{eq:cartan02}), and (\ref{eq:cartan03}) are
satisfied. For the classical case,
$\Sigma$ is the phase space and ${\cal L}$ is the set of subsets of the
phase space. The Cartan map is the
identity.
Of course, in the two examples, quantum and classical, the states 
correspond with
atoms of the lattice of properties. By
means of two axioms we regain this property for the general axiomatic
situation. Since, as we mentioned
already, the structure of a state-property system is derived from the
operational aspects of the axiomatic
approach, these two axioms should be considered as the first two axioms of
the new axiomatic approach where
states are not identified {\it a priori} with the atoms of the 
property lattice.
Let us give these two axioms.

\subsection{The First Two Axioms: State Determination and Atomisticity}

A physical entity $S$ is described by its state-property system $(\Sigma,
{\cal L}, \kappa)$, where $\Sigma$
is a set, its elements representing the states of $S$, ${\cal L}$ is a
complete lattice, its elements
representing the properties of $S$, and $\kappa$ is a map from ${\cal L}$
to ${\cal P}(\Sigma)$, satisfying
(\ref{eq:cartan01}), (\ref{eq:cartan02}), and (\ref{eq:cartan03}), and
expressing the physical situation:
``The property $a \in {\cal L}$ is actual if the entity $S$ is in state $p
\in
\Sigma$" by $p \in \kappa(a)$. This is the structure that we derive from
only operational aspects of the
axiomatic approach. The first axiom that we introduce consists in demanding
that a state is determined by
the set of properties that are actual in this state.

\begin{axiom} [State Determination]
For $p, q \in \Sigma$ such that
\be
\bigwedge_{p \in \kappa(a)}a = \bigwedge_{q \in \kappa(b)}b
\ee
we have $p =
q$.
\end{axiom}
We remark that in \cite{Aerts1981,Aerts1982,Piron1976,Piron1990} this axiom
is considered to be
satisfied {\it a priori}. The second axiom consists in demanding that 
the states
can be considered as atoms of the
property lattice.

\begin{axiom} [Atomisticity]
For $p \in \Sigma$ we have that
\be
\bigwedge_{p \in \kappa(a)}a
\ee
is an atom of ${\cal L}$.
\end{axiom}
Obviously these two axioms are satisfied for the two examples ${\cal
P}(\Omega)$ and ${\cal P}({\cal H})$ that we
considered.

\subsection{The Third Axiom: Orthocomplementation}
For the third axiom it is already very difficult to give a complete physical
interpretation. This third axiom introduces the structure of
an orthocomplementation for the lattice of properties. At first sight the
orthocomplementation could be seen as a structure that
plays a similar role for properties as the negation in logic plays for
propositions. But that is not a very careful
way of looking at things. We cannot go into the details of the attempts
that have been made to interpret the orthocomplementation
in a physical way, and refer to
\cite{Piron1976,Piron1990,Aerts1981,Aerts1982,Aerts1983a} for those that
are interested in this
problem. Also in
\cite{Valckenborgh2000,Valckenborgh2001,DDH2002,Aerts2002a} the problem is
considered in
depth.

\begin{axiom} [Orthocomplementation]
The lattice ${\cal L}$ of properties of the physical entity under study is
orthocomplemented. This means that there exists a
function $' :{\cal L} \rightarrow {\cal L}$ such that for $a, b \in {\cal
L}$ we have:
\bea
(a')' &=& a \\
a < b &\Rightarrow& b' < a' \\
a \wedge a' = 0\ &{\rm and}&\ a \vee a' = I
\eea
\end{axiom}
For ${\cal P}(\Omega)$ the orthocomplement of a subset is given by the
complement of this subset, and for ${\cal P}({\cal H})$
the orthocomplement of a closed subspace is given by the
subspace orthogonal to this closed subspace.

\subsection{The Fourth and Fifth Axiom: The Covering Law and Weak Modularity}

The next two axioms are called the covering law and weak modularity. There
is no obvious physical interpretation for
them. They have been put forward mainly because they are satisfied in the
lattice of closed subspaces of a complex Hilbert space.
These two axioms are what we have called the `extra conditions' when we talked
about Piron's representation theorem in the
introduction of this section.

\begin{axiom} [Covering Law]
The lattice ${\cal L}$ of properties of the physical entity under study
satisfies the covering law. This means that for $a, x \in
{\cal L}$ and $p \in \Sigma$ we have:
\be
a < x < a \vee p \Rightarrow x = a\ {\rm or}\ x = a \vee p
\ee
\end{axiom}

\begin{axiom} [Weak Modularity]
The orthocomplemented lattice ${\cal L}$ of properties of the physical
entity under study is weakly modular. This means that for
$a , b \in {\cal L}$ we have:
\be
a < b \Rightarrow (b \wedge a') \vee a = b
\ee
\end{axiom}
These are the five axioms of standard quantum axiomatics. It can be
shown that both axioms, the covering law and weak
modularity, are satisfied for the two examples ${\cal P}(\Omega)$ and
${\cal P}({\cal H})$ \cite{Piron1964,Piron1976}.

The two examples that we have mentioned show that both classical entities
and quantum entities can be described by the common structure
of a complete atomistic orthocomplemented lattice that satisfies the
covering law and is weakly modular. Now we have to consider the
converse, namely how this structure leads us to classical physics and to
quantum physics.

%%%%%%%%%%%%%%%%%%%%%%%%%%%%%%%%%%%%%%%%%%%%%%%%%%%%%%%%%%%%%%%%%%%%%%%%%%%%%%%

\section{The Representation Theorem}

First we show how the classical and nonclassical parts can be extracted
from the general structure, and second we show how
the nonclassical parts can be represented by so-called generalized 
Hilbert spaces.

\subsection{The Classical and Nonclassical Parts}

Since both examples ${\cal P}(\Omega)$ and ${\cal P}({\cal H})$ satisfy the
five axioms, it is clear that a theory where the
five axioms are satisfied can give rise to a classical theory, as well as
to a quantum
theory. It is possible to filter out the classical part by introducing the
notions of classical property and classical state.

\bd [Classical Property]
Suppose that $(\Sigma, {\cal L}, \kappa)$ is the state property system
representing a physical
entity, satisfying axioms 1, 2 and 3. We say that a property $a \in {\cal
L}$ is a classical property if
for all $p \in \Sigma$ we have
\be
p \in \kappa(a) \ {\rm or}\ p \in \kappa(a')
\ee
The set of all classical properties we denote by ${\cal C}$.
\ed
Again considering our two examples, it is easy to see that for the quantum
case, hence for ${\cal L} = {\cal P}({\cal H})$, we
have no nontrivial classical properties. Indeed, for any closed subspace $A
\in {\cal H}$, different from $0$ and ${\cal
H}$, we have rays of ${\cal H}$ that are neither contained in $A$ nor
contained in $A'$. These are exactly the rays that
correspond to states that are superposition states of states in $A$ and
states in $A'$. It is the superposition principle in
standard quantum mechanics that makes that the only classical properties of
a quantum entity are the trivial ones, represented
by $0$ and ${\cal H}$. It can also easily be seen that for the case of a
classical entity, described by ${\cal P}(\Omega)$, all
the properties are classical properties. Indeed, consider an arbitrary
property $A \in {\cal P}(\Omega)$, then for any singleton
$\{p\} \in \Sigma$ representing a state of the classical entity, we have
$\{p\} \subset A$ or $\{p\} \subset A'$, since $A'$ is
the set theoretical complement of $A$.

\bd [Classical State] \label{def:classicalstate}
Suppose that $(\Sigma, {\cal L}, \kappa)$ is the state property system of a
physical entity satisfying axioms
1, 2 and 3. For $p \in \Sigma$ we introduce
\bea
\omega(p) &=& \bigwedge_{p \in \kappa(a), a \in {\cal C}}a \\
\kappa_c(a) &=& \{\omega(p)\ \vert \ p \in \kappa(a)\}
\eea
and call $\omega(p)$ the classical state of the physical entity 
whenever it is in
a state $p \in \Sigma$, and $\kappa_c$ the classical Cartan map. The set of
all classical states
will be denoted by $\Omega$.
\ed

\bd [Classical State Property System]
Suppose that $(\Sigma, {\cal L}, \kappa)$ is the state property system of a
physical entity satisfying axioms
1, 2 and 3.  The classical state property system corresponding with
$(\Sigma, {\cal L}, \kappa)$ is $(\Omega,
{\cal C}, \kappa_c)$.
\ed
Let us look at our two examples. For the quantum case, with ${\cal L}
= {\cal P}({\cal H})$, we have only two classical
properties, namely $0$ and ${\cal H}$. This means that there is only one
classical state, namely ${\cal H}$. It is the classical
state that corresponds to `considering the quantum entity under study' and
the state does not specify anything more than that.
For the classical case, every state is a classical state.

It can be proven that $\kappa_c: {\cal C} \rightarrow {\cal P}(\Omega)$ is an
isomorphism\cite{Aerts1981,Aerts1983a}.
This means that if we filter out the classical part and limit the
description of our general physical entity to its classical
properties and classical states, the description becomes a standard
classical physical description.

Let us filter out the nonclassical part.

\bd [Nonclassical Part] \label{def:nonclassicalcomponents}
Suppose that $(\Sigma, {\cal L}, \kappa)$ is the state property system of a
physical entity satisfying axioms
1, 2 and 3. For $\omega \in \Omega$ we introduce
\bea
{\cal L}_\omega &=& \{a\ \vert a < \omega, a \in {\cal L}\} \\
\Sigma_\omega &=& \{p\ \vert p \in \kappa(\omega), p \in \Sigma\} \\
\kappa_\omega(a) &=& \kappa(a)\ {\rm for}\ a \in {\cal L}_\omega
\eea
and we call $(\Sigma_\omega, {\cal L}_\omega, \kappa_\omega)$ the nonclassical
components of $(\Sigma, {\cal L}, \kappa)$.
\ed
For the quantum case, hence ${\cal L} = {\cal P}({\cal H})$, we have only
one classical state ${\cal H}$, and obviously ${\cal
L}_{\cal H} = {\cal L}$. Similarly we have $\Sigma_{\cal H} = \Sigma$. This
means that the only nonclassical
component is $(\Sigma, {\cal L}, \kappa)$ itself.
For the classical case, since all properties are classical properties and
all states are classical states, we have ${\cal
L}_\omega = \{0,\omega\}$, which is the trivial lattice, containing only its
minimal and maximal element, and $\Sigma_\omega =
\{\omega\}$. This means that the nonclassical components are all trivial.

For the general situation of a physical entity described by $(\Sigma, {\cal
L}, \kappa)$ it can be shown that
${\cal L}_\omega$ contains no classical properties with respect to
$\Sigma_\omega$ except $0$ and
$\omega$, the minimal and maximal element of ${\cal
L}_\omega$, and that if $(\Sigma, {\cal L}, \kappa)$ satisfies axioms 1, 2,
3, 4, and 5, then
also
$(\Sigma_\omega, {\cal L}_\omega, \kappa_\omega)\
\forall \omega \in \Omega$ satisfy axioms 1, 2, 3, 4 and 5 (see
\cite{Aerts1981,Aerts1983a}).

We remark that if axioms 1, 2 and 3 are satisfied we can identify a state $p
\in \Sigma$ with the element of
the lattice of properties ${\cal L}$ given by:
\be
s(p) = \bigwedge_{p \in \kappa(a), a \in {\cal L}}a
\ee
which is an atom of ${\cal L}$. More precisely, it is not difficult to verify
that, under the assumption of axioms 1 and 2, $s : \Sigma \to \Sigma_{\cal
L}$ is a well-defined mapping that is one-to-one and onto, 
$\Sigma_{\cal L}$ being
the collection of all atoms in
$\cal L$. Moreover,
$p
\in
\kappa(a)$ iff
$s(p)
< a$. We can call
$s(p)$ the property state corresponding to $p$ and define
\be
\Sigma' = \{s(p)\ \vert\ p \in \Sigma\} \label{eq:stateproperty}
\ee
the set of state properties. It is easy to verify that if we introduce
\be
\kappa': {\cal L} \rightarrow {\cal P}(\Sigma')
\ee
where
\be
\kappa'(a) = \{s(p)\ \vert\ p \in \kappa(a)\} \label{eq:statepropertycartan}
\ee
that
\be
(\Sigma', {\cal L}, \kappa') \cong (\Sigma, {\cal L}, \kappa)
\ee
when axioms 1, 2 and 3 are satisfied.

To see in
more detail in which way the classical and nonclassical parts are
structured within the lattice ${\cal L}$, we
make use of this isomorphism and introduce the direct union of a set of
complete, atomistic
orthocomplemented lattices, making use of this identification.

\bd [Direct Union] \label{def:directunion}
Consider a set $\{{\cal L}_\omega\ \vert \omega \in \Omega\}$ of complete,
atomistic orthocomplemented lattices. The direct union
$\directunion_{\omega \in \Omega}{\cal L}_\omega$ of these lattices
consists of the sequences $a = (a_\omega)_\omega$, such
that\bea(a_\omega)_\omega < (b
_\omega)_\omega &\Leftrightarrow& a_\omega < b_\omega\ \forall \omega \in
\Omega \\
(a_\omega)_\omega \wedge (b_\omega)_\omega &=& (a_\omega \wedge
b_\omega)_\omega  \\
(a_\omega)_\omega \vee (b_\omega)_\omega &=& (a_\omega \vee b_\omega)_\omega \\
(a_\omega)_\omega' &=& (a_\omega')_\omega
\eea
The atoms of $\directunion_{\omega \in \Omega}{\cal L}_\omega$ are of the
form $(a_\omega)_\omega$ where $a_{\omega_1} = p$ for
some
$\omega_1$ and $p \in \Sigma_{\omega_1}$, and $a_\omega = 0$ for $\omega
\not= \omega_1$.
\ed
It can be proven that if ${\cal L}_\omega$ are complete, atomistic,
orthocomplemented lattices, then also $\directunion_{\omega
\in
\Omega}{\cal L}_\omega$ is a complete, atomistic, orthocomplemented lattice
(see \cite{Aerts1981,Aerts1983a}).
The structure of direct union of complete, atomistic, orthocomplemented
lattices makes it possible to define
the direct union of state property systems in the case axioms 1, 2, and 3
are satisfied.

\bd [Direct Union of State Property Systems] Consider a set of state
property systems $(\Sigma_\omega, {\cal
L}_\omega, \kappa_\omega)$, where ${\cal L}_\omega$ are complete,
atomistic, orthocomplemented lattices and
for each $\omega$ we have that $\Sigma_\omega$ is the set of atoms of
${\cal L}_\omega$. The direct union
$\directunion_\omega (\Sigma_\omega, {\cal L}_\omega, \kappa_\omega)$ of
these state property systems is the
state property system $(\cup_\omega\Sigma_\omega, \directunion_\omega{\cal
L}_\omega,
\directunion_\omega\kappa_\omega)$, where $\cup_\omega\Sigma_\omega$ is the
disjoint union of the sets
$\Sigma_\omega$, $\directunion_\omega{\cal L}_\omega$ is the direct union
of the lattices ${\cal L}_\omega$,
and
\be
\directunion_\omega\kappa_\omega ((a_\omega)_\omega) =
\cup_\omega\kappa_\omega(a_\omega)
\ee
\ed
The first part of
a fundamental representation theorem can now be stated. For this part it is
sufficient that axioms 1, 2 and 3 are satisfied.

\begin{theorem} [Representation Theorem: Part 1] \label{theor:representation01}
We consider a physical entity described by its state property system
$(\Sigma, {\cal L}, \kappa)$. Suppose
that axioms 1, 2 and 3 are satisfied. Then
\be
(\Sigma, {\cal L}, \kappa) \cong \directunion_{\omega \in
\Omega}(\Sigma'_\omega, {\cal L}_\omega,
\kappa'_\omega)
\ee
where $\Omega$ is the set of classical states  of $(\Sigma, {\cal L},
\kappa)$ (see definition
\ref{def:classicalstate}), $\Sigma'_\omega$ is the set
of state properties,
$\kappa'_\omega$ the corresponding Cartan map, (see
(\ref{eq:stateproperty}) and
(\ref{eq:statepropertycartan})), and
${\cal L}_\omega$ the lattice of properties (see definition
\ref{def:nonclassicalcomponents}) of the
nonclassical component
$(\Sigma_\omega, {\cal L}_\omega,
\kappa_\omega)$.
If axioms 4 and 5 are
satisfied for $(\Sigma, {\cal L}, \kappa)$, then they are also satisfied for
$(\Sigma'_\omega, {\cal L}_\omega,
\kappa'_\omega)$ for all $\omega \in
\Omega$.
\end{theorem}
Proof: see \cite{Aerts1981,Aerts1983a}

\subsection{Further Representation of the Nonclassical Components}

 From the previous section follows that if axioms 1, 2, 3, 4 and 5 are
satisfied we can write the state property system $(\Sigma, {\cal L},
\kappa)$ of the physical entity under
study as the direct union
$\directunion_{\omega \in \Omega}(\Sigma'_\omega, {\cal L}_\omega,
\kappa'_\omega)$ over its classical
state space $\Omega$ of its nonclassical components
$(\Sigma'_\omega, {\cal L}_\omega,
\kappa'_\omega)$, and that each of these nonclassical components also
satisfies axiom 1, 2, 3, 4 and 5. Additionally for each
one of these nonclassical components
$(\Sigma'_\omega, {\cal L}_\omega,
\kappa'_\omega)$ no classical properties except $0$ and $\omega$ exist. It
is for these nonclassical components that a
further representation theorem can be proven such that a vector space
structure emerges for each one of the nonclassical
components. To do this we rely on the original representation theorem that
Piron proved in \cite{Piron1964}.

\begin{theorem} [Representation Theorem: Part 2] \label{theor:representation02}
Consider the same situation as in theorem \ref{theor:representation01},
with additionally axiom 4 and 5 satisfied. For
each nonclassical component
$(\Sigma'_\omega, {\cal L}_\omega,
\kappa'_\omega)$, of which the lattice ${\cal L}_\omega$ has at least four
orthogonal states\footnote{Two
states $p, q \in \Sigma_\omega$ are orthogonal if there
exists
$a \in {\cal L}_\omega$ such that
$p < a$ and $q < a'$.}, there exists a vector space
$V_\omega$, over a division ring $K_\omega$, with an involution of
$K_\omega$, which means a function
\be
^*: K_\omega \rightarrow K_\omega
\ee
such that for $k, l \in K_\omega$ we have:
\bea
(k^*)^* &=& k \\
(k \cdot l)^* &=& l^* \cdot k^*
\eea
and an Hermitian product on $V_\omega$, which means a function
\be
\langle\ ,\ \rangle: V_\omega \times V_\omega \rightarrow K_\omega
\ee
such that for
$x, y, z \in V_\omega$ and $k \in K_\omega$ we have:
\bea
\langle x + ky, z \rangle &=& \langle x, z \rangle + k \langle x, y \rangle \\
\langle x, y \rangle^* &=& \langle y, x \rangle \\
\langle x, x \rangle = 0 &\Leftrightarrow& x = 0
\eea
and such that for $M \subset V_\omega$ we have:
\be
M^\perp + (M^\perp)^\perp = V_\omega
\ee
where $M^\perp = \{y\ \vert y \in V_\omega,
\langle y, x
\rangle = 0, \forall x \in M\}$. Such a vector space is called a
generalized Hilbert space or an
orthomodular vector space. And we have that:
\be
(\Sigma'_\omega, {\cal L}_\omega, \kappa'_\omega) \cong ({\cal R}(V), {\cal
L}(V), \nu)
\ee
where ${\cal R}(V)$ is the set of rays of $V$, ${\cal L}(V)$ is the set of
biorthogonally closed subspaces (subspaces
that are equal to their biorthogonal) of $V$, and $\nu$ makes correspond
with each such biorthogonal
subspace the set of rays that are contained in it.
\end{theorem}
Proof: See \cite{Piron1964,Piron1976}.

%%%%%%%%%%%%%%%%%%%%%%%%%%%%%%%%%%%%%%%%%%%%%%%%%%%%%%%%%%%%%%%%%%%%%%%%%%%%%%%%%

\section{The Two Failing Axioms of Standard Quantum Mechanics}

We have introduced all that is necessary to be able to put forward the
theorem that has been proved
regarding the failure of standard quantum mechanics for the description of
the joint entity consisting of two separated quantum
entities \cite{Aerts1981,Aerts1982}. Let us first explain what is
meant by separated physical entities.

\subsection{What Are Separated Physical Entities?} 
\label{sec:separatedentities}
We consider the situation of a physical entity $S$ that consists of two
physical entities $S_1$ and $S_2$. The definition of
`separated' that has been used in
\cite{Aerts1981,Aerts1982} is the following. Suppose that we consider two
experiments $e_1$ and $e_2$ that can be performed
respectively on the entity $S_1$ and on the entity $S_2$, such that the
joint experiments $e_1 \times e_2$ can be
performed on the joint entity $S$ consisting of $S_1$ and $S_2$. We say
that experiments $e_1$ and $e_2$ are separated
experiments whenever for an arbitrary state $p$ of $S$ we have that 
$(x_1, x_2)$
is a possible outcome for experiment $e_1 \times e_2$
if and only if $x_1$ is a possible outcome for $e_1$ and $x_2$ is a
possible outcome for $e_2$. We say that $S_1$ and
$S_2$ are separated entities if and only if all the experiments $e_1$ on
$S_1$ are separated from the experiments $e_2$ on $S_2$.

Let us remark that $S_1$ and $S_2$ being separated does not mean that there
is no interaction between $S_1$ and
$S_2$. Most entities in the macroscopic world are separated entities. Let
us consider some examples to make this clear.

The
earth and the moon, for example, are separated entities. Indeed, consider
any experiment $e_1$ that can be performed on the
physical entity earth (for example measuring its position), and any
experiment $e_2$ that can be performed on the physical entity
moon (for example measuring its velocity). The joint experiment $e_1 \times
e_2$ consists of performing $e_1$ and $e_2$ together
on the joint entity of earth and moon (measuring the position of the earth
and the velocity of the moon at once). Obviously
the requirement of separation is satisfied. The pair $(x_1,x_2)$
(position of the earth and velocity of the moon) is a possible
outcome for $e_1
\times e_2$ if and only if
$x_1$ (position of the earth) is a possible outcome of
$e_1$ and $x_2$ (velocity of the moon) is a possible outcome of $e_2$. This
is what we mean when we say that the earth has
position
$x_1$ and the moon velocity $x_2$ at once. Clearly this is independent of
whether there is an interaction, the gravitational
interaction in this case, between the earth and the moon.

It is not easy to find an example of two physical entities that are not
separated in the macroscopic world,
because usually nonseparated entities are described as one entity and not
as two. In earlier work we have
given examples of nonseparated macroscopic entities
\cite{Aerts1982b,Aerts1984,Aerts1988}. The example of connected vessels of
water is a good example to give an intuitive idea of what nonseparation means.

Consider two vessels $V_1$ and $V_2$ each
containing 10 liters of water. The vessels are connected by a tube, which
means that they form a connected set of vessels. Also
the tube contains some water, but this does not play any role for what we
want to show. Experiment $e_1$ consists of taking out
water of vessel $V_1$ by a siphon, and measuring the amount of water that
comes out. We give the outcome $x_1$ if the
amount of water coming out is greater than 10 liters. Experiment $e_2$
consists of doing exactly the same on vessel $V_2$. We
give outcome $x_2$ to $e_2$ if the amount of water coming out is greater
than 10 liters. The joint experiment $e_1
\times e_2$ consists of performing $e_1$ and $e_2$ together on the joint
entity of the two connected vessels of water. Because of
the connection, and the physical principles that govern connected vessels,
for $e_1$ and for $e_2$ performed alone we find 20 liters
of water coming out. This means that $x_1$ is a possible (even certain)
outcome for $e_1$ and $x_2$ is a possible (also certain)
outcome for
$e_2$. If we perform the joint experiment
$e_1 \times e_2$ the following happens. If there is more than 10 liters
coming out of vessel $V_1$ there is less than 10 liters
coming out of vessel $V_2$ and if there is more than 10 liters coming out
of vessel $V_2$ there is less than 10 liters coming out
of vessel $V_1$. This means that $(x_1, x_2)$ is not a possible outcome for
the joint experiment $e_1 \times e_2$. Hence $e_1$
and $e_2$ are nonseparated experiments and as a consequence $V_1$ and $V_2$
are nonseparated entities.

The nonseparated entities that we find in the macroscopic world are
entities that are very similar to the
connected vessels of water. There must be an ontological connection between
the two entities, and that is also the reason that
usually the joint entity will be treated as one entity again. A connection
through dynamic interaction, as it is the case
between the earth and the moon, interacting by gravitation, leaves the
entities separated.

For quantum entities it can be shown that only when the joint entity of two
quantum entities contains entangled states
the entities are nonseparated quantum entities. It can be proven
\cite{Aerts1982b,Aerts1984,Aerts1988} that experiments are
separated if and only if they do not violate Bell's inequalities. All this
has been explored and investigated in many ways, and
several papers have been published on the matter
\cite{Aerts1982b,Aerts1984,Aerts1988,Aerts1990,ABG2000}. Interesting
consequences for the Einstein Podolsky Rosen paradox and the violation of
Bell's inequalities have been
investigated \cite{Aerts1985,Christiaens2002}.

\subsection{The Separated Quantum Entities Theorem}

We are ready now to state the theorem about the impossibility for
standard quantum mechanics to describe separated
quantum entities \cite{Aerts1981,Aerts1982}.

\begin{theorem} [Separated Quantum Entities Theorem] \label{theor:theorem03}
Suppose that $S$ is a physical entity consisting of two separated physical
entities $S_1$ and $S_2$. Let us suppose that axiom 1,
2 and 3 are satisfied and call $(\Sigma, {\cal L}, \kappa)$ the state
property system
describing $S$, and
$(\Sigma_1, {\cal L}_1, \kappa_1)$ and $(\Sigma_2, {\cal L}_2, \kappa_2)$
the state property systems
describing
$S_1$ and $S_2$.

\noindent
If the fourth axiom is satisfied, namely the covering law, then one of the
two entities $S_1$ or $S_2$ is a classical
entity, in the sense that one of the two state property systems $(\Sigma_1,
{\cal L}_1, \kappa_1)$ or
$(\Sigma_2, {\cal L}_2, \kappa_2)$ contains only classical states and
classical properties.

\noindent
If the fifth axiom is satisfied, namely weak modularity, then one of the
two entities $S_1$ or $S_2$ is a classical
entity, in the sense that one of the two state property systems $(\Sigma_1,
{\cal L}_1, \kappa_1)$ or
$(\Sigma_2, {\cal L}_2, \kappa_2)$ contains only classical states and
classical properties.
\end{theorem}
Proof: see \cite{Aerts1981,Aerts1982}

\medskip
\noindent
The theorem proves that two separated quantum entities cannot be described
by standard quantum mechanics. A classical entity
that is separated from a quantum entity and two separated classical
entities do not cause any problem, but two separated quantum
entities need a structure where neither the covering law nor weak
modularity are satisfied.

One of the possible ways out is that there would not exist separated
quantum entities in nature. This would mean that all
quantum entities are entangled in some way or another. If this is true,
perhaps the standard formalism could be saved. Let
us remark that even standard quantum mechanics presupposes the existence of
separated quantum entities. Indeed, if we
describe one quantum entity by means of the standard formalism, we take one
Hilbert space to represent the states of this entity.
In this sense we suppose the rest of the universe to be separated from this
one quantum entity. If not, we would have to modify
the description and consider two Hilbert spaces, one for the entity and one
for the rest of the universe, and the states would be
entangled states of the states of the entity and the states of the rest of
the universe. But, this would mean that the one
quantum entity that we considered is never in a well-defined state.
It would mean that the only possibility that remains is to describe the
whole universe at once by using one huge Hilbert space. It goes without
saying that such an approach will lead to many other
problems. For example, if this one Hilbert space has to describe the whole
universe, will it also contain itself, as a
description, because as a description, a human activity, it is part of the
whole universe. Another, more down to earth problem
is, that in this one Hilbert space of the whole universe also all classical
macroscopical entities have to be described. But
classical entities are not described by a Hilbert space, as we have seen in
section \ref{sec:quantumaxiomatics}. If the
hypothesis that we can only describe the whole universe at once is correct,
it would anyhow be more plausible that the theory
that does deliver such a description would be the direct union structure of
different Hilbert spaces. But if this is the case, we anyhow
are already using a more general theory than standard quantum mechanics. So
we can as well use the still slightly more general theory,
where axioms 4 and 5 are not satisfied, and make the description of
separated quantum entities possible.

All this convinces us that the shortcoming of standard quantum mechanics to
be able to describe separated quantum entities is
really a shortcoming of the mathematical formalism used by standard quantum
mechanics, and more notably of the vector space
structure of the Hilbert space used in standard quantum mechanics.

\subsection{Operational Foundation of Quantum Axiomatics}

To be able to explain the conceptual steps that are made to prove theorem
\ref{theor:theorem03} we have to explain how the concept of
`separated' is expressed in the quantum axiomatics that we introduced in
section \ref{sec:quantumaxiomatics}. Separated entities are
defined by means of separated experiments. In the quantum axiomatics of
section \ref{sec:quantumaxiomatics} we do not talk about
experiments, which means that there is still a link that is missing. This
link is made by what is called the operational foundation of
the quantum axiomatic lattice formalism. Within this operational foundation
a property of the entity under study is defined by the
equivalence class of all experiments that test this property. We will not
explain the details of this operational foundation, because
some subtle matters are involved, and refer to
\cite{Piron1976,Aerts1981,Aerts1982} for these details. What we need to
close the circle
in this article is the fact that, making use of the operational
foundations, it is possible to introduce `separated properties' as
properties that are defined by equivalence classes of separated experiments.

\subsection{The Separated Quantum Entities Theorem Bis}

Theorem \ref{theor:theorem03} can then be reformulated
completely in the language of the axiomatic quantum formalism that we
introduced in section \ref{sec:quantumaxiomatics} in the following
way:

\begin{theorem} [Separated Quantum Entities Theorem Bis]
Suppose that we consider the compound entity $S$ that consists of two physical
entities
$S_1$ and $S_2$. Suppose that axioms 1, 2 and 3 are satisfied for
$S$, $S_1$ and $S_2$. Suppose that each property of
$S_1$ is `separated' from each property of $S_2$. If axiom
4 is satisfied for
$S$, then one of the two entities $S_1$ or
$S_2$ contains only
classical properties and classical states, and hence $S_1$ or $S_2$ is a
classical entity. If axiom 5 is satisfied for $S$, then one of the two
entities $S_1$ or
$S_2$ contains only
classical properties and classical states, and hence $S_1$ or $S_2$ is a
classical entity.
\end{theorem}

\subsection{Linearity at Stake}
If the covering law is not satisfied for the lattice ${\cal L}$ that
describes the compound entity consisting of two separated quantum
entities,
then this lattice cannot be represented into a vector space. This
means that the
superposition principle will not be valid for $S$. In
standard quantum mechanics, situations have been encountered where the
superposition principle is not valid, and one refers to these
situations as `the presence of superselection rules'. For example the
property `charge' for a microparticle entails such a
superselection rule. There are no superpositions of states with different
charge. It has always been possible to incorporate
superselection rules into the standard formalism by demanding that there
should be no superpositions between different sectors of the
Hilbert space. The reason that this could be done for superselection rules
as the ones that arise from the property charge, is because
the states that correspond to different values of a physical quantity are
always orthogonal. This has made it possible to circumvent the
problem by considering different orthogonal sectors of a common Hilbert space,
and not allowing superpositions between states of
different sectors. It can be shown that the superselection rules that arise
from the situation of separated quantum entities correspond
to states that are not orthogonal, which means that the traditional way of
avoiding the problem cannot work. In \cite{Aerts1981,Aerts1982} explicit
examples of states that are
separated by a super selection rule are given. Also in \cite{AV2002} we
give examples of such states.

\subsection{Some Subtle Aspects of the Separated Quantum Entities Theorem}

The `Separated Quantum Entities Theorem' that was proved in
\cite{Aerts1981,Aerts1982} was correctly
criticized by Cattaneo and Nistic\'o \cite{CN1990}. As we mentioned already,
the proof in
\cite{Aerts1981,Aerts1982} is made by introducing separated experiments,
where separated is defined as
explained in section \ref{sec:separatedentities}. Then separated properties
are defined as properties that
are tested by separated experiments, and once the property lattice of the
joint entity is constructed in
this way, the theorem can be proven. The whole construction in
\cite{Aerts1981,Aerts1982} is built by
starting with only yes/no-experiments, hence experiments that have only two
possible outcomes. The reason
that the construction in \cite{Aerts1981,Aerts1982} is made by means of
yes/no-experiments has a purely
historical origin. The version of operational quantum axiomatics elaborated in
Geneva, where one of the authors
was working when proving the separated quantum entities
theorem, was a version where only
yes/no-experiments are considered as basic operational concepts. There did
exist at that time versions of
operational quantum axiomatics that incorporated right from the start 
experiments
with any number of possible
outcomes as basic operational concepts, as for example the approach elaborated
by Randall and Foulis
\cite{RF1976,RF1978,FR1981}. Cattaneo and Nistic\'o proved in \cite{CN1990}
that, by considering only
yes/no-experiments as an operational basis for the construction of the
property lattice of the compound entity consisting of
separated entities, some of the possible experiments that can be performed
on this compound entity are
overlooked. It could well be that the experiments that had been overlooked
in the construction of
\cite{Aerts1981,Aerts1982} were exactly the ones that, once added, would
give rise to additional
properties and make the lattice of properties satisfy again axiom 4 and 5.
That is the reason that
Cattaneo and Nistic\'o state explicitly in their article \cite{CN1990} that
they do not question the
mathematical argument of the proof, but rather its
operational basis. This was indeed a
serious critique that had been pondered carefully. Although the
author involved in
this matter remembers clearly that he was convinced then that the lattice
of properties would not
change by means of the addition of the lacking experiments indicated by
Cattaneo and Nistic\'o, and that
his theorem remained valid, there did not seem an easy way to prove
this. The only way out was to
redo the construction but now starting with experiments with any
number of outcomes as basic
operational concepts. This is done in \cite{Aerts1994}, and indeed, the
separated quantum entities theorem
can also be proved with this operational basis. This means that in
\cite{Aerts1994} the critique of
Cattaneo and Nistico has been answered, and the result is that the theorem
remains valid. The construction
presented in \cite{Aerts1994} is however much less transparent than the
original one to be found in
\cite{Aerts1981,Aerts1982}. That is the reason why it is interesting to
analyze the most simple
of all situations of such a compound entity, the one consisting of
two separated spin
1/2 objects. This is exactly what we will do in
\cite{AV2002}. On this simple example it is easy to go through the full
construction of the lattice ${\cal L}$ and its set of states $\Sigma$, such
that we can see how fundamentally different it is from a
structure that would entail a vector space type of linearity. Note that
since the separated quantum
entities theorem is a no-go theorem, also the simple example of
\cite{AV2002} contains a proof of the
no-go aspect of the original theorem.

%%%%%%%%%%%%%%%%%%%%%%%%%%%%%%%%%%%%%%%%%%%%%%%%%%%%%%%%%%%%%%%%%%%%%%%%%%%%%%%%%

\section{Attempts and Perspectives for Solutions}
In this section we mention briefly what are the attempts that have
meanwhile taken place to find a solution
to the problem that we have considered in this paper.

If we consider the aspect of the Separated Quantum Entities Theorem where
an explicit construction of the
lattice of properties and set of states of the compound entity consisting
of separated subentities
is made, then the theorem proves that this construction cannot be made
within standard quantum mechanics,
from which follows that standard quantum mechanics cannot describe
separated quantum entities. Of course,
in its profound logical form the Separated Quantum Entities Theorem is a
no-go theorem, which means that
also some of the other hypotheses that are used to prove the theorem can be
false and hence also at the
origin of the problem. Research, which partially took place even before the
Separated Quantum Entities
Theorem, and partially afterwards, gives us some valuable extra 
information about
what are the possible directions
that could be explored to `solve' the problem connected with the Separated
Quantum Entities Theorem.

\subsection{Earlier Research on the Compound Entity Problem}
At the end of the seventies, one of the authors
studied the problem of the description of compound entities in quantum
axiomatics, but this time
staying within the quantum axiomatic framework where each considered entity
is described by a complex
Hilbert space, as in standard quantum mechanics \cite{AD1978a,AD1978b}.
This means that the quantum
axiomatic framework was only used to give an alternative but equivalent
description of standard quantum
mechanics, because even then the quantum axiomatic framework makes it
possible to translate physical
requirements in relation with the situation of a compound physical entity
consisting of two quantum
mechanical subentities. The main aim of this research on the problem was
to find back the tensor product
procedure of standard quantum mechanics for the description of the compound
entity, but this time not as
an {\it ad hoc} procedure, which it is in standard quantum mechanics, but from
physically interpretable
requirements. For these  requirements, some so-called `coupling
conditions' were put forward.

\begin{theorem} \label{ther:couplingtheorem}
We describe quantum entities $S_1$, $S_2$ and $S$, respectively by their
Hilbert space lattices (sets of
closed subspaces of the Hilbert space),
${\cal L}({\cal H}_1)$, ${\cal L}({\cal H}_2)$ and ${\cal L}({\cal H})$,
and by their Hilbert space state spaces (sets of rays of the
Hilbert spaces) $\Sigma({\cal H}_1)$, $\Sigma({\cal H}_2)$ and
$\Sigma({\cal H})$. Suppose that dim ${\cal
H}_1 > 2$ and dim ${\cal H}_2 >2$. Suppose that $h_1$, $h_2$ are functions:
\bea
h_1: {\cal L}({\cal H}_1) &\rightarrow& {\cal L}({\cal H}) \\
h_2: {\cal L}({\cal H}_2) &\rightarrow& {\cal L}({\cal H})
\eea
such that for all $A_1, B_1, C_1, (A_1^i)_i \in {\cal L}({\cal H}_1)$, $A_2,
B_2, C_2, (A_2^j)_j \in {\cal
L}({\cal H}_2)$,
$p_1
\in \Sigma({\cal H}_1)$ and $p_2 \in \Sigma({\cal H}_2)$ we have
\bea
A_1 \subset B_1 &\Rightarrow& h_1(A_1) \subset h_1(B_1)
\label{eq:coupling01} \\
A_2 \subset B_2 &\Rightarrow& h_2(A_2) \subset h_2(B_2)
\label{eq:coupling02} \\
h_1(\vee_iA_1^i)) &=& \vee_ih_1(A_1^i) \label{eq:coupling03} \\
h_2(\vee_iA_2^i)) &=& \vee_ih_2(A_2^i) \label{eq:coupling04} \\
h_1({\cal H}_1) &=& h_2({\cal H}_2) = {\cal H} \label{eq:coupling05} \\
h_1(C_1) &\leftrightarrow& h_2(C_2)
\label{eq:coupling06} \\ h_1(p_1) &\wedge& h_2(p_2) \in \Sigma({\cal H})
\label{eq:coupling07}
\eea
where $\leftrightarrow$ is the symbol for `compatible', then ${\cal
P}({\cal H})$ is canonically isomorphic
to
${\cal P}({\cal H}_1
\otimes {\cal H}_2)$ or to
${\cal P}({\cal H}_1 \otimes {\cal H}_2^*)$.
\end{theorem}
Proof: see \cite{AD1978a,AD1978b}

\medskip
\noindent
The conditions (\ref{eq:coupling01}),
(\ref{eq:coupling02}),(\ref{eq:coupling03}),(\ref{eq:coupling04}),
(\ref{eq:coupling05}),(\ref{eq:coupling06}) and (\ref{eq:coupling07}) are
called the `coupling conditions'
in \cite{AD1978a,AD1978b}. The physical interpretation for the different
conditions is quite
straightforward. Conditions (\ref{eq:coupling01}),
(\ref{eq:coupling02}),(\ref{eq:coupling03}),(\ref{eq:coupling04}) and
(\ref{eq:coupling05}) mean that the functions $h_1$ and $h_2$ are morphisms
of the lattice structure.
Hence they express that $S_1$ and $S_2$ can be recognized as subentities
of $S$. Condition
(\ref{eq:coupling06}) expresses that properties of $S_1$ are compatible
with properties of $S_2$, and
condition (\ref{eq:coupling07}) expresses that when $S_1$ and $S_2$ are in
certain states, then $S$ is in a state
uniquely determined by these states of $S_1$ and $S_2$.

When the article \cite{AD1978b} was written, the authors aimed at giving a
physical justification for
using the tensor product in standard quantum mechanics for the description
of the compound entity consisting of two
quantum entities. The theorem succeeds well in doing so. There is however
one remarkable aspect of the
theorem. Two possible solutions appear and they are not canonically
isomorphic. This means that for the
category of Hilbert space lattices and their morphisms none of the two
solutions can be a categorical
product, because than the two solutions should be canonically isomorphic.
This is amazing, because
one would expect that if one moves to the mathematical level that
corresponds well with the physics,
which should be the Hilbert space lattice rather than the Hilbert space
itself, one would find one of
the categorical products to correspond to what is needed for the
description of the compound entity.
Let us remark that the theorem shows that if the
Hilbert spaces would all be real Hilbert spaces instead of
complex, there is
only one solution, in which case it could be a categorical product. Of
course, it is well known that the
complex numbers play an essential role in quantum mechanics, such that the
two solutions do represent
different entities.

\subsection{Investigating Further a Categorical Approach}
Becoming aware of the fact that no categorical solution can be inferred
from theorem
\ref{ther:couplingtheorem}, it became interesting to look straightforwardly
for a categorical
construction. This is what was done in \cite{Aerts1984}. A categorical
product, more specifically a
co-product, can be constructed, but it gives a structure that is very
different from what one gets in
standard quantum mechanics (the tensor product of Hilbert spaces), and from
what one gets from theorem
\ref{ther:couplingtheorem}. A theorem that is very similar to the Separated
Quantum Entities Theorem can be
proven for the co-product. Again two of the axioms of traditional quantum
axiomatics are never satisfied
for the compound entity of two quantum entities if we would describe this
compound entity by means of the
co-product, except when one of the subentities is a trivial entity, with
a lattice of properties containing
only $0$ and $I$ \cite{Aerts1984}. One of the failing axioms is again the
covering law, which means that
also here, if we choose to use the co-product instead of the separated
product to describe the compound
entity, linearity is gone. We cannot go more into detail on this matter in
this paper, but refer to
\cite{Valckenborgh2000,Valckenborgh2001} where the situation of the three
products, the separated product,
the co-product and the Hilbert space tensor product, is studied in detail
by one of the authors.

\subsection{The Problem of Pure States and Mixed States}
  From what we have explained in the foregoing, the situation is such that
(1) the compound entity consisting of two
separated quantum entities cannot be described by standard quantum
mechanics, and (2)
there remains an unsolved
problem in relation with the description of the compound entity of two (not
necessarily separated) quantum
entities, in the sense that traditional quantum mechanics only knows the
tensor product of Hilbert spaces
procedure, but this procedure cannot be fitted into an operational scheme
at the axiomatic level. These
results seem to indicate strongly that standard quantum mechanics must be
generalized in the sense that a
mathematical formalism should be worked out where the covering law is
dropped, and hence linearity is
lost. More recently however another possibility has come to the surface.
Since also this possibility is
relevant for the whole of the book where this article appears, we want to
explain it briefly.

The
Separated Quantum Entities Theorem of \cite{Aerts1982} and the Co-Product
Theorem of \cite{Aerts1984} are
in essence no-go theorems. And although both theorems give a very strong
argument in favor of the view
that standard quantum mechanics should be generalized by dropping the
covering law and hence loosing
linearity, we have to be careful. The most profound conclusion that has to
be drawn from any no-go theorem
is that at least one of the hypotheses that is used to prove the theorem is
false. This means that the
situation may even be worse, namely that not only the covering law (and
weak modularity) should be
dropped, but that there is even another, perhaps more important, hypothesis
false in standard quantum
mechanics. Of course, normally one would start to elaborate a
generalization by dropping the least
possible number of hypotheses necessary. But our research shows that even if we
drop the covering law (and weak
modularity), and construct then the co-product, we still do not find a
satisfactory way to describe the
compound entity of two quantum entities. Moreover, as we mentioned, the
co-product structure is so
different from the tensor product of Hilbert spaces structure used in
standard quantum mechanics in a more
or less fruitful way, that it might well be that we are not looking at the
right category. This type of
reflections and other ones have led one of the authors to consider
the following possibility:
{\it perhaps we should
reconsider the way in which pure states are described in quantum mechanics
by means of rays of the Hilbert
space. And more concretely, perhaps also density operators, that normally
are interpreted as only
describing mixed states, represent pure states as well as mixed states}.
This idea has been considered and
introduced in \cite{Aerts1999} and the physical and philosophical
situation connected with it has been
analyzed in \cite{Aerts2000}. The quantum formalism where one
allows density operators to
represent pure states has been called `extended quantum mechanics' in
\cite{Aerts1999}.

It is clear that this
conceptual change will not solve
all of the problems, for example, the fact that separated quantum entities
cannot be described by extended
standard quantum mechanics is still true. This means that also extended
quantum mechanics shall have to be
formulated by means of a structure that is different from the complex
Hilbert space that is used in
standard quantum mechanics. But something is also gained that might make
that the mathematical change that
is needed under extended quantum mechanics is less drastic than the one
that is needed under standard
quantum mechanics. We can see this by noting that for extended quantum
mechanics the state of a compound
entity, whether it is a ray state or a density operator state, is always a
product state of states of the
subentities. This comes from the `mathematical' fact that any density
operator in the tensor product
Hilbert space is a product of density operators in the component Hilbert
spaces. This means that there is
more hope that a categorical construction for the lattice of properties and
set of states of an extended
quantum mechanics would give rise to a co-product that is closer to the
tensor product of Hilbert spaces
structure that is now used in standard quantum mechanics. We cannot say
much more about this now, because
we did not have the time to investigate sufficiently the operational and
categorical structures that go
along with extended quantum mechanics. We plan to engage in this
investigation in the future. What we can
see immediately is that if density operators also represent pure states,
the axiom of atomicity will not
be fulfilled for the pure states that arise from density operators.

\end{document}